# Effective Clustering Algorithms for Gene Expression Data

T.Chandrasekhar          K.Thangavel          E.Elayaraja

Department of Computer Science, Periyar University,
Salem – 636 011, Tamil Nadu, India.

## ABSTRACT
Microarrays are made it possible to simultaneously monitor the expression profiles of thousands of genes under various experimental conditions. Identification of co-expressed genes and coherent patterns is the central goal in microarray or gene expression data analysis and is an important task in Bioinformatics research. In this paper, K-Means algorithm hybridised with Cluster Centre Initialization Algorithm (CCIA) is proposed Gene Expression Data. The proposed algorithm overcomes the drawbacks of specifying the number of clusters in the K-Means methods. Experimental analysis shows that the proposed method performs well on gene Expression Data when compare with the traditional K- Means clustering and Silhouette Coefficients cluster measure.

## 1. INTRODUCTION
Most Data mining algorithms developed for microarray gene expression data deal with the problem of clustering. Cluster analysis of gene expression data has proved to be a useful tool for identifying co-expressed genes. DNA microarrays are emerged as the leading technology to measure gene expression levels primarily, because of their high throughput. Results from these experiments are usually presented in the form of a data matrix in which rows represent genes and columns represent conditions or samples [12]. Each entry in the matrix is a measure of the expression level of a particular gene under a specific condition. Analysis of these data sets reveals genes of unknown functions and the discovery of functional relationships between genes [19]. Co-expressed genes can be grouped into clusters based on their expression patterns of gene. Clustering can be performed based on genes and samples. In gene based clustering, the genes are treated as the objects, while the samples are the features. In sample based clustering, the samples can be partitioned into homogeneous groups where the genes are regarded as features and the samples as objects. [19].

Gene Data sets are pre-processed using normalization and discretization methods. An attribute is normalized by scaling its values so that they fall within a small specified range, such as -1.0 to 1.0, or 0.0 to 1.0. Normalization is particularly useful for classification algorithms involving neural networks, or distance measurements such as nearest-neighbor classification and clustering. Normalizing the input values for each attribute measured in the training tuples will help to speed up the learning phase. For distance-based methods, normalization helps prevent attributes with initially large ranges from outweighing attributes with initially smaller ranges. There are many methods for data normalization as min-max normalization, z-score normalization, and normalization by decimal scaling. Many machine learning techniques can only be applied to data sets composed entirely of nominal variables but a very large proportion of real data sets include continuous variables. One solution to this problem is to partition numeric variables into a number of sub-ranges and treat each such sub-range as a category. This process of partitioning continuous variables into categories is usually termed as discrerization.

Cluster analysis is a one of the primary data analysis tools in data mining.The most popular clustering algorithms in microarray gene expression analysis are Hierarchical Clustering, K-Means clustering [3], and SOM [9]. Of these K-Means clustering is a very simple and fast efficient one; it was developed by Mac Queen [7]. The K-Means algorithm is effective in producing clusters for many practical applications. But the computational complexity of the original K-Means algorithm is very high, especially for large Data sets. The K-Means clustering algorithm is a partitioning clustering method that separates data into K groups. For real life problems, the suitable number of clusters cannot be predicted. To overcome the above drawback the current research focused on developing the clustering algorithms without giving the initial number of clusters [2, 5].

This paper is organized as follows. Section 2 presents an overview of various preprocessing techniques. Section 3 describes the unsupervised Clustering algorithms. Section 4 describes performance of Experimental analysis and discussion. Section 5 conclusion and future work.

## 2. PRE PROCESSING TECHNIQUES
The purpose of clustering gene expression data is to reveal the natural structure inherent data and extracting useful information from noisy data. So, pre-processing of the data is an essential part in any data mining process. Some of the methods require the input data to be discrete, taken rough sets, clustering and association rules [4]. That is why a new task in pre-processing is needed as normalization and unsupervised discretization.

### 2.1 Discretization Method-I
The gene expression data is normalized to have mean 0 and standard deviation 1. Gene Expression Data having a low variance across conditions as well as Z-Score normalization methods are used. The standard score is

$$z = \frac{x - \mu}{\sigma}$$

Where:  $x$ is a raw score to be standardized
μ is the mean of the population

σ is the standard deviation of the population.

Discretization is then performed on this normalized expression data. The discretization is done as follows [17]

i. The discretized value of gene $g_i$ at condition, $t_1$ (i.e., the first condition)

$$\xi_{g_i,t_1} = \begin{cases} 1 & if\ \varepsilon_{g_i,t_1} > 0 \\ 0 & if\ \varepsilon_{g_i,t_1} = 0 \\ -1 & if\ \varepsilon_{g_i,t_1} < 0 \end{cases}$$

ii. The discretized values of gene $g_i$ at conditions $t_j$ ($j = 1,..(T − 1)$) i.e., at the rest of the conditions $(T − \{t_1\})$

$$\xi_{g_i,t_{j+1}} = \begin{cases} 1 & if\ \varepsilon_{g_i,t_j} < \varepsilon_{g_i,t_{j+1}} \\ 0 & if\ \varepsilon_{g_i,t_j} = \varepsilon_{g_i,t_{j+1}} \\ -1 & if\ \varepsilon_{g_i,t_j} > \varepsilon_{g_i,t_{j+1}} \end{cases}$$

Here $\xi g_i, t_j$ is the discretized value of gene $g_i$ at conditions $t_j$ ($j = 1, 2, ..., T − 1$). The gene expression values are gene $g_i$ and condition $t_j$ is given by $g_i, t_j$. Compute first condition $t_1$, is treated as a special case and its discretized value is directly based on $g_i, t_1$ i.e., the expression value at condition $t_1$. For the rest of the conditions, the discretized value is calculated by comparing its expression value with that of the previous value. This helps in finding whether the gene is up 1 or down -1 regulated at that particular condition. Each gene will now have a regulation pattern of 0, 1, and -1 across the conditions or time points. This pattern is represented as a string.

## 2.2 Discretization Method-II
The gene expression data is min-max normalization by setting min 0 and max1. Min-max normalization performs a linear transformation on the original data. Suppose that $min_A$ and $max_A$ *are* the minimum and maximum values of an attribute, *A*. Min-max normalization maps a value, *v*, of *A* to $v_0$ in the range [new_$min_A$, new_$max_A$] by computing

$$v' = \frac{v - min_A}{max_A - min_A}(new\_max_A - new\_min_A) + new\_min_A$$

Min-max normalization preserves the relationships among the original data values. It will encounter an "out-of-bounds" error if a future input case for normalization falls outside of the original data range for *A*. After the normalization of discretized value of gene $g_i$ at condition, $t_j$ is given

$$\xi_{g_i,t_j} = \begin{cases} 1 & if\ \varepsilon_{g_i,t_j}\ 0\ to\ 0.25 \\ 2 & if\ \varepsilon_{g_i,t_j}\ 0.25\ to\ 0.5 \\ 3 & if\ \varepsilon_{g_i,t_j}\ 0.5\ to\ 0.75 \\ 4 & if\ \varepsilon_{g_i,t_j}\ 0.75\ to\ 1 \end{cases}$$

So, each gene will now have a regulation pattern of 1, 2, 3 and 4 across the conditions or time points. This pattern is represented as a string.

## 2.3 Discretization Method-III
The gene expression data is default normalization using by MATLAB 9.0 in Bioinformatics Toolbox™ functions. In this Normalize microarray data scales the values in each column of microarray data by dividing by the mean column intensity. In this function as follow

**xNorm = manorm(x)**

Where: x - Microarray data , Enter a vector or matrix.
xNorm - Normalized microarray data.
Manorm - Microarray normalization.

After this normalization method, to follow the discretization.

Equal width interval method [5] is to discretize data into k equally sized bins, where k is a parameter provided by users. Given a variable v, the upper bound and lower bound of v are $v_{max}$ and $v_{min}$, respectively. Then, this method use formula below to compute the size of bin

$$\lambda = \frac{v_{max} - v_{min}}{k}$$

Hence, for $i^{th}$ bin, the bin boundary is [bi-1; bi], bi = $b_0$+iλ, where $b_0$ = $v_{min}$, i = 1, 2, ... , k- 1.

Equal width interval method can be applied to each feature indendently. Since this method doesn't class information, thus it is an unsupervised discretization method. In the problem of dense regions discovery, given a data matrix X, $v_{ij}$ represents the value of entry xij, so we can also apply Equal width interval method to discretize X. In such cases, we define as each binning (k) has various range numbers. (Ex. bin 1 as 1, bin 2 as 2 ...) So, each gene will now have a regulation pattern of 1, 2, 3 ... across the conditions or time points. This pattern is represented as a string.

## 2.4 Discretization Method-IV
The gene expression data is default normalization using by MATLAB 9.0 in Bioinformatics Toolbox™ functions. Normalize rows of matrix is normr(M) normalizes the columns of M to a length of 1.

**N = normr(x)**

Here x - gene expression data, N - Normalized data.

After that normalization then we distinguish between balanced expression, under expression and over expression of genes. Since no further information about the cut-off levels is available, the gene expression profiles are discretized according to the following rules. (1) All negative values are considered as -1. (2) All positive values are considered as 1. (3) All other balanced values are considered as 0. So, each gene will now have a regulation pattern of -1, 1, and 0 across the conditions or time points. This pattern is represented as a string.

## 3. UNSUPERVISED CLUSTERING ALGORITHMS
Clustering is one of the unsupervised methods; each cluster is a collection of objects which are similar to each other and are dissimilar to the objects belonging to other clusters. The similarity mostly is measured with distance: two or more objects belong to the same cluster if they are close according to a given distance [2]. Sometimes similarity is measured referring to a concept representing a cluster. Two or more objects belong to the same cluster if it defines a concept common to all these objects. In other words, objects are grouped according to their fit to a descriptive concept.

## 3.1 K- Means Clustering

The main objective in cluster analysis is to group objects that are similar in one cluster and separate objects that are dissimilar by assigning them to different clusters. One of the most popular clustering methods is K-Means clustering algorithm [3, 9, 12,]. It is classifies objects to a pre-defined number of clusters, which is given by the user (assume *K* clusters). The idea is to choose random cluster centres, one for each cluster. These centres are preferred to be as far as possible from each other. In this algorithm mostly Euclidean distance is used to find distance between data points and centroids [6, 13]. The Euclidean distance between two multi-dimensional data points $X = (x_1, x_2, x_3, ..., x_m)$ and $Y = (y_1, y_2, y_3, ..., y_m)$ is described as follows

$$D(X, Y) = \sqrt{(x_1 - y_1)^2 + (x_2 - y_2)^2 + \cdots + (x_M - y_M)^2}$$

The K-Means method aims to minimize the sum of squared distances between all points and the cluster centre. This procedure consists of the following steps, as described below.

**Algorithm 1**: K-Means clustering algorithm [18]

**Require**: $D = \{d_1, d_2, d_3, ..., d_n\}$ // Set of n data points.
 K - Number of desired clusters
**Ensure**: A set of K clusters.

**Steps:**
1. Arbitrarily choose $k$ data points from *D* as initial centroids;
2. **Repeat**
   Assign each point $d_i$ to the cluster which has the closest centroid;
   Calculate the new mean for each cluster;
   **Until** convergence criteria is met.

Though the K-Means algorithm is simple, it has some drawbacks of quality of the final clustering, since it highly depends on the arbitrary selection of the initial centroids [1, 14].

## 3.2 Cluster Centre Initialization Algorithm (CCIA)

Performance of iterative clustering algorithms which converges to numerous local minima depends highly on initial cluster centers. Generally initial cluster centers are selected randomly. In this section, the cluster centre initialization algorithm is studied to improve the performance of the K-Means algorithm.

**Algorithm 2**: Cluster Centre Initialization Algorithm [18]

**Input**: $D = \{d_1, d_2, ..., d_n\}$ // set of *n* data items
 K // Number of desired clusters
**Output**: A set of K initial centroids.
**Steps**:

1. Set m = 1;
2. Compute the distance between each data point and all other data points in the set D;
3. Find the closest pair of data points from the set D and form a data point set $A_m (1 \leq m \leq K)$ which contains these two data points, Delete these two data points from the set D;
4. Find the data point in D that is closest to the data point set $A_m$, Add it to $A_m$ and delete it from D;
5. Repeat step 4 until the number of data points in $A_m$ reaches 0.75 * (n/K);
6. If m < K, then m = m+1, find another pair of data points from D between which the distance is the shortest, form another data-point set $A_m$ and delete them from D, Go to step 4;
7. For each data point set $A_m (1 \leq m \leq K)$ find the arithmetic Mean of the vectors of data points in $A_m$, these means will be the initial centroids.

## 4. EXPERIMENTAL ANALYSIS AND DISCUSSION

In this section, we describe the data sets used to analyse the methods studied in sections 2 and 3, which are arranged for the listed in Table.1, number of features/genes are in column wise, and number of items/samples are in row wise.

### Serum data
This data set is described and used in [10]. It can be downloaded from: http://www.sciencemag.org/feature/data/984559.shl and corresponds to the selection of 517 genes whose expression varies in response to serum concentration inhuman fibroblasts [11].

### Yeast data
This data set is downloaded from Gene Expression Omnibus-databases. The Yeast cell cycle dataset contains 2884 genes and 17 conditions. To avoid distortion or biases arising from the presence of missing values in the data matrix we removal all the genes that had any missing value. This step results in a matrix of size 2882 * 17.

### Simulated data
It is downloaded from http://www.igbmc.ustrasbg.fr/projets/fcm/y3c.txt. The set contains 300 Genes [3].

### Leukemia data set
It is downloaded from the website: http://datam.i2r.a-star.edu.sg/datasets/krbd/. The set contains 7129*34.

### 4.1. Comparative and Performance analysis
The K-Means, CCIA with K-Means used to clustering in the all data sets after the section 2 the distance measure used here is the Euclidean distance. To access the quality of the clusters, we used the silhouette measure proposed by Rousseeuw [11, 15, 16]. In this K-Means method the Initial centroid values taken 12 then run as 10 times running to K-Means as initial random seed value choose, clusters data into 12 groups as Table 1 and Fig 1.

**Table 1. Comparative analysis of Gene Expression Data in K-Means and CCIA with K-Means**

| Data Set | Methods | Actual Data set | Discretization Method-I | Discretization Method-II | Discretization Method-III | Discretization Method-IV |
|---|---|---|---|---|---|---|
| Serum | K-Means | 0.3297 | 0.2282 | 0.3761 | 0.3146 | 0.2198 |
| | CCIA with K-Means | 0.5159 | 0.4909 | 0.3170 | 0.5318 | 0.3456 |
| Yeast | K-Means | 0.3189 | 0.2994 | 0.2778 | 0.2119 | 0.1912 |
| | CCIA with K-Means | 0.5196 | 0.4801 | 0.3130 | 0.5216 | 0.3893 |
| Simulated | K-Means | 0.3197 | 0.2896 | 0.2135 | 0.2125 | 0.2248 |
| | CCIA with K-Means | 0.5186 | 0.4256 | 0.3469 | 0.5193 | 0.3913 |
| Leukemia | K-Means | 0.3474 | 0.3092 | 0.2751 | 0.2897 | 0.2823 |
| | CCIA with K-Means | 0.5497 | 0.4973 | 0.3797 | 0.5597 | 0.4127 |

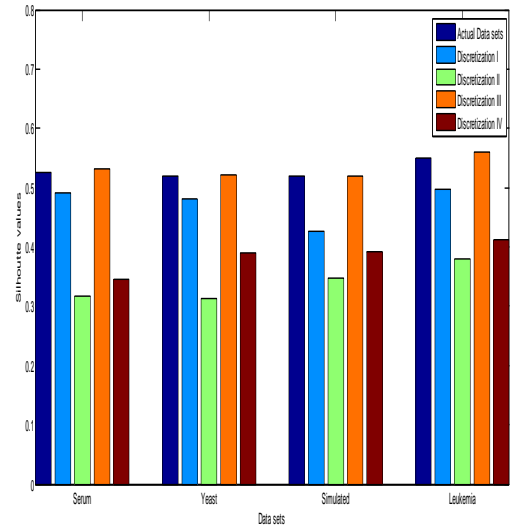

**Fig.2 Clusters Performance comparison chart for all data sets in CCIA with K-Means**

## 6. CONCLUSION
In this work, various preprocessing methods and unsupervised clustering are studied to apply and avoid too many redundant or missing values in Microarray gene expression data to improve the quality of cluster. These methods are used to get minimum number of random gene data sets and then we use K-Means clustering technique to improve the quality of clusters. One of the demerits of K-Means algorithm is random selection of initial seed point of desired clusters. This was overcome with CCIA for finding the initial centroids to avoid the random selection of initial values. Therefore, the CCIA algorithm is not depending upon any choice of the number of cluster and automatic evaluation of initial seed centroids and it produces different better results. Both the algorithms were tested with gene expression data and analysis the performance of cluster values using silhouette measurement. Therefore, finding solution to various pre-processing of normalization and discretization, select the different centroids as clusters seed points, various measures are used to improve the cluster performance is our future endeavour.

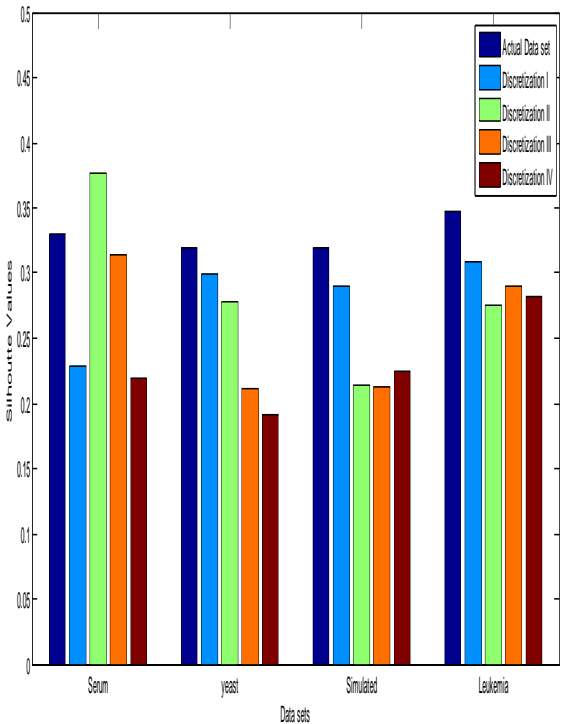

**Fig.1 Clusters Performance comparison chart for all data sets in K-Means**

In this CCIA with K-Means method the initial K value is taken 12, then the seed points are automatically select and execute 10 times to clusters data into 12 groups as Table 1 and Fig2. Applying CCIA with K-Means in actual data set that the results is better than the K-Means results. In the same way, when the normalization and discretization is made, we get the optimal result in CCIA with K-Means. Applying normalization with CCIA in an actual data set, it produces the best result. Especially the 3$^{rd}$ discretized method of Microarray Normalization method combined with CCIA and K-Means algorithms produces the best result for Serum, Yeast, Simulated and Leukemia data sets.